\documentclass[a4paper]{article}
\usepackage{hyperref}
\usepackage[T1]{fontenc} 
\usepackage[utf8]{inputenc} 
\usepackage{amsmath}         
\usepackage{amssymb}
\usepackage{amsthm}
\usepackage{bbm}             
\usepackage{bm}              
\usepackage[mathscr]{eucal}  
\usepackage{blindtext}

\newcommand{\ist}[1]{\overset{\footnotesize(\ref{#1})}{=}}
\newcommand{\ca}[1]{\overset{\footnotesize(\ref{#1})}{\approx}}

\title{Energy as a measure for the elapse of time}
\date{\today}
\author{Manfried Faber\footnote{faber@kph.tuwien.ac.at}\\[3mm]
Atominstitut\\
Technische Universität Wien\\
 Operngasse 9, 1040 Vienna, Austria}
\begin{document}
\maketitle
\begin{abstract}
Clocks in different heights or with different velocities run with different speeds. For global positioning systems these effects are much too large to be ignored. Nevertheless, in classical and quantum mechanics we get high accuracy using a ``universal'' time scale, not depending on altitudes and velocities. One may ask how this is possible. The answer to this question we may get from the observation that in classical and quantum mechanics time and energy are canonically conjugate variables. We argue that the mentioned modifications of the time scale by relativistic effects are taken into account in the notion of energy. On the basis of the experimental results and the laws of special relativity we argue that we should consider energy as measure for the elapse of time.
\end{abstract}


\section{Introduction}
The notion of energy is of central importance in modern culture and technology. It allows us to calculate how we can substitute man power by engines and thus facilitate daily life. We trade energy and we even ascribe prices to the different types of energy. We talk of fossil fuels, oil and gas, and renewable energies, wind and water power. In these cases we do not mean the types of energy but the types of fuels.

In physics we define different types of energy, potential, kinetic and thermal energy and explain the conservation of energy. We know that stable systems are located at a minimum of potential energy. The mathematical expression for the energy, the Hamilton function $H$, has become a central notion of many branches of physics. From $H$ we can derive the stability of systems and the time evolution of unstable systems. If we know the particles and fields, thus the degrees of freedom of a system and the expression for the energy, we have the important information to do successful calculations.

Nevertheless, we can ask, what energy really is. We realise that energy is strongly related to time. Energy is conserved, if the Hamilton function is invariant against translations of time. In classical mechanics, the evolution of a system is given by the temporal derivatives of the generalized coordinates $q_i$ and momenta $p_i$, which are expressed in terms of the partial derivatives of the Hamilton function $H(q_k,p_k)$ according to Hamilton's equations~\cite{Gosson:2018aa}
\begin{equation}
\dot q_i=\frac{\partial H(q_k,p_k)}{\partial p_i},\quad
\dot p_i=-\frac{\partial H(q_k,p_k)}{\partial q_i}.
\end{equation}
The evolution of a quantum mechanical system is described by the time-evolution operator~\cite{Messiah:1995aa}
\begin{equation}\label{ZeitEntOp}
\exp\{-\mathrm iHt/\hbar\},
\end{equation}
where $H$ is the Hamiltonian of the system. In submicroscopic physics the product of energy and time, the action $S$, is measured in units of Planck's constant $h=2\pi\hbar=6,626\cdot10^{-34}~$Js, one of the fundamental constants of nature.

\section{Experimental results}
The relation of energy and time shows up even more directly in experiments. The precise measurements of time, which are available today with atomic clocks, demonstrate that clocks of the same type are running faster, if they are moved to regions of higher potential energy or if the velocity is decreased~\cite{Hafeleaa:1972,Hafeleab:1972}. Today measurements are enormously precise~\cite{Chou24092010}. For a difference in height $\Delta h=0.33~$m a relative difference of frequency $\nu$
\begin{equation}
\frac{\Delta\nu}{\nu}=(4,1\pm1,6)\cdot10^{-17}
\end{equation}
was measured. This value agrees with the relative increase of the energy
\begin{equation}\label{GravEneZunahme}
\frac{m_0g\,\Delta h}{m_0c^2}=3,6\cdot10^{-17}
\end{equation}
of a mass $m_0$ in the gravitational field with acceleration $g$. Time dilations for velocities of less than 10 meters per second where reported in the above mentioned article agreeing with the predictions of special relativity.

\section{Theoretical reasoning}
These experimental results may help us better understand the notion of energy. Despite the fact that clocks at different altitudes run with different speeds, we relate our (universal) time scale to Coordinated Universal Time (UTC), based on International Atomic Time (TAI), a weighted average of the time kept at present by the proper time~\footnote{The proper time is the time measured along a timelike world line in four-dimensional spacetime. Thus, clocks by definition measure the proper time.} of over 400 atomic clocks. In order to avoid errors, using this time in classical and quantum mechanics, we have learned to account for the different speeds of clocks in the product $Et$. We use a universal time scale and adjust the factor energy, and in this way we take the variations of proper time into account. Therefore, an increase in the  potential energy
\begin{equation}\label{EneZunahme}
E=m_0c^2\;\to\;E=m_0c^2+m_0g\Delta h
\end{equation}
indicates that the system is in a region, where the proper time $\tau$ runs faster, see Eq.~(\ref{GravEneZunahme}),
\begin{equation}
\mathrm d\tau\;\to\;\mathrm d\tau^\prime
=\mathrm d\tau(1+\frac{g\,\Delta h}{c^2}),
\end{equation}
or more precise for varying gravitation acceleration $\vec a(\vec r)$ along a path $\mathcal C$
\begin{equation}
\mathrm d\tau\;\to\;\mathrm d\tau^\prime=\mathrm d\tau
\big(1+\frac{1}{c^2}\int_{\mathcal C}\vec a(\vec r)\mathrm d\vec r\big).
\end{equation}
If we insert the acceleration
\begin{equation}
\vec a=-\frac{GM}{r^2}\vec e_r
\end{equation}
in the gravitational field of the earth of mass $M$ we get
\begin{equation}\label{PotEinfl}
\mathrm d\tau^\prime
=\mathrm d\tau\left[1+\frac{GM}{c^2}\left(\frac{1}{r_1}
-\frac{1}{r_2}\right)\right]\quad\textrm{with}\quad\Delta h=r_2-r_1,\;
g=\frac{GM}{r^2}.
\end{equation}

In quantum mechanics, where the states are represented by the wave functions $\psi(\vec r,t)$, the time evolution can be expanded in a Taylor series which can be formally summed up
\begin{equation}\label{TaylorZeitEntw}
\psi(\vec r,t)=\sum_{k=0}^\infty\frac{t^k\partial_t^k}{k!}\,\psi(\vec r,t)
\Big|_{t=0}=\mathrm e^{t\,\partial_t}\psi(\vec r,t)\Big|_{t=0}.
\end{equation}
As $\partial_t$ is an antihermitean operator we get real eigenvalues for the hermitean operator $\mathrm i\partial_t$. Using universal time $t$, we have to take into account the different speeds of time in the operator $\mathrm i\partial_t$. According to the time dependent Schrödinger equation
\begin{equation}\label{ZASchroGl}
\mathrm i\hbar\partial_t\psi=H\psi,
\end{equation}
the temporal derivative of the wave function is given by the Hamiltonian $H$ acting on that function. With Eq.~(\ref{ZASchroGl}), the time evolution operators~(\ref{ZeitEntOp}) and (\ref{TaylorZeitEntw}) become identical. $H$ takes into account that the energy is proportional to the speed of time, as measured with precise clocks.

In the path integral formulation of quantum mechanics~\cite{Feynman:1948ur,Feynman:1965aa,Ramond:1981pw,Kleinert:2004ev} the time evolution of a free particle in the comoving frame is described by the transition amplitude
\begin{equation}\label{PfadFaktor}
\mathrm e^{-\mathrm i\tau\,m_0c^2/\hbar}.
\end{equation}
To get the general expression we have to perform a Lorentz transformation from the comoving frame, we call it $\mathcal S^\prime$ with the coordinates $t^\prime=\tau$ and $\vec r^\prime=(x^\prime,y^\prime,z^\prime)=0$, to the laboratory frame $\mathcal S$ moving with the velocity $-\vec v$ relative to $\mathcal S^\prime$,
\begin{equation}\label{LorTrafoTau}
\tau=t^\prime=\gamma(t-\frac{\vec v\vec r}{c^2}),\quad \vec r^\prime=0\quad\textrm{with}\quad
\gamma=\frac{1}{\sqrt{1-\beta^2}},\quad\beta=\frac{v}{c}.
\end{equation}
Multiplying by $m_0c^2$ we get
\begin{equation}\label{LorTrafo}
m_0c^2\tau=\underbrace{m_0c^2\gamma}_{E}t-\underbrace{m_0\gamma\vec v}_{\vec p}\vec r.
\end{equation}
Introducing energy $E$ and momentum $\vec p$
\begin{equation}\label{DefEp}
E:=\gamma m_0c^2,\quad\vec p:=\gamma m_0\vec v
\end{equation}
we shift the dependence on the speed of time to these velocity depending quantities. Thus, the transition amplitude~(\ref{PfadFaktor}) transforms from the comoving to the laboratory frame
\begin{equation}\label{ZeitEntwLabor}
\mathrm e^{-\mathrm i\tau\,m_0c^2/\hbar}
\ist{LorTrafo}\mathrm e^{\mathrm i(\vec p\vec r-Et)/\hbar}.
\end{equation}
The propagation of a free particle is described by the action $S=\vec p\vec r-Et$, where energy and momentum are related by
\begin{equation}\label{EneImpBez}
E^2\ist{DefEp}m_0^2c^4+\vec p^{\,2}c^2
\end{equation}
and in the  non relativistic approximation (v<<c) by
\begin{equation}\label{EApprox}
E\ca{EneImpBez}m_0c^2+\frac{\vec p^{\,2}}{2m_0}.
\end{equation}

If the velocity $\vec v$ of a mass $m_0$ relative to an inertial frame $\mathcal S$ depends on time, we attribute this to an interaction. From the kinematical laws of special relativity we get the relation for the force $\vec F$~\cite{Goldstein:2002aa}
\begin{equation}\label{EnergieErh}
m_0c^2\mathrm d\gamma=\vec F\mathrm d\vec r.
\end{equation}
If the force $\vec F$ is an integrable quantity, it can be expressed by the gradient of a potential $V(\vec r)$
\begin{equation}\label{DefPot}
\vec F=-\vec\nabla V(\vec r).
\end{equation}
Integrating Eq.~(\ref{EnergieErh}) we get the well-known relativistic expression for the energy $E$
\begin{equation}\label{GesamtENichtRel}
m_0c^2\;\to\;E=\gamma m_0c^2+V\ist{EApprox}m_0c^2+T+V\quad\textrm{with}\quad
T\approx\frac{\vec p^{\,2}}{2m_0}.
\end{equation}
This equation does also include the case of the gravitational field, see Eqs.~(\ref{EneZunahme})-(\ref{PotEinfl}), if we write
\begin{equation}\label{GraviV}
V(r)=-G\frac{Mm_0}{r}.
\end{equation}
Thus, taking into account both effects, moving frames and the effects of conservative forces, the time evolution (\ref{ZeitEntwLabor}) generalises to
\begin{equation}\label{WirkungBeideBeitr}
S=\int L\mathrm dt,\quad L=\vec p\,\dot{\vec r}-H,\quad H=\gamma m_0c^2+V.
\end{equation}

In the case of fields the Lagrangian $L$ can be written as an integral over a Lagrange density $\mathcal L$ with $L=\int\mathcal L\,\mathrm d^3r$. The transition amplitude for quantum fields is then given by
\begin{equation}\label{BoltzFak}
\mathrm e^{\mathrm iS/\hbar},\quad S=\frac{1}{c}\int \mathrm d^4x\,\mathcal L,\quad
x=(ct,\vec r).
\end{equation}
The fact that in a quantum system the action $S$ is measured in natural units of $\hbar$, reflects the property of path integrals that quantum fields are not restricted to the minimum of action. The action can fluctuate in the order of $\hbar$ around these minima~\cite{Feynman:1948ur,Feynman:1965aa,Ramond:1981pw,Kleinert:2004ev}.

Every particle has its Compton time $\hbar/(m_0c^2)$, which also appears in the Schrödinger equation in natural units
\begin{equation}\label{scaleSchroGl}
\mathrm i\frac{\hbar}{m_0c^2}\partial_t\psi(\vec r,t)
=\left[1-\frac{1}{2}\left(\frac{\hbar}{m_0c}\right)^2\Delta
+\frac{V}{m_0c^2}\right]\psi(\vec r,t).
\end{equation}
This variation of the time scale with the rest mass of the particle is in agreement with the claim that we are using in calculations a universal time scale. If the particle would annihilate with an antiparticle, we say in the energy language that its rest energy $m_0c^2$ could be transformed to work on another mass or to an increase of its velocity. The ratio of proper time to universal time for this other particle would change accordingly. The rest energy  of a particle has the ability to change the scale of the proper time of other particles. Every subsystem of a composite system has therefore its own time scale given by its Compton time. As the temporal derivative in Eq.~(\ref{TaylorZeitEntw}) of  a composite system appears in the exponent, the temporal derivatives of the subsystems are summed up.

Momenta are canonically conjugate to the coordinates. The gradient $\vec\nabla$ is an antihermitean operator and performs a shift of the coordinates, in the same way as $\partial_t$ shifts the time, see Eq.~(\ref{TaylorZeitEntw}). Under Lorentz transformation the quantum mechanical length scale, which for the particle at rest is the Compton wavelength  $\hbar/(m_0c)$, is modified according to Eqs.~(\ref{LorTrafo}) and (\ref{DefEp}) with the factor $\gamma$ and indicates that in our calculations we are also using universal length scales.

General relativity introduces coordinate lines in a four-dimensional pseudo-Rie\-mannian manifold, coordinates are labels to specify events in 3+1-dimensional spacetime. Gravity is not regarded as a force but a manifestation of the curvature of spacetime. The concept of proper time is directly used without transition to the concept of gravitational forces and gravitational energy. Since gravitational effects are taken into account in the metric, it is impossible to define a generally covariant energy-momentum tensor for the gravitational field~\cite{Padmanabhan:2010zzb}. The energy-momentum tensor takes into account only the strong, electromagnetic and weak interactions as sources of curvature and determines the time- and length scales. The transition to a direct description of these interactions in terms of the degrees of freedom of space- and time is not done. The unification of gravitation and particle physics is still missing. The largest part of the present community of physicists in searching for a quantisation of gravity. One could also think of another type of unification by a geometrisation of particle physics. A step in this direction concerning electromagnetism, the only long-range interaction besides gravity, was done in Ref.~\cite{Faber:1999ia,Faber:2002nw,Faber:2012zz,Faber:2014bxa} on the classical level.

\section{Conclusions}
This paper addresses the conceptual issue how classical and quantum mechanics can account for the difference in the speed of clocks depending on their velocity or the strength of an external gravitational field. The precise measurements with atomic clocks have impressively verified the exact predictions of Special and General Relativity. Classical and quantum mechanics, on the other hand, are using a ``universal'' time scale, not depending on the velocities of the clocks or the strength of the gravitational field. Energy is introduced as quantity canonically conjugate to time. The product of energy and time is measured in units of one of the fundamental constants of nature, the quantum of action, Planck's constant $\hbar$. We argue in this article that during the history of physics we learned to construct the expressions for the energy in such a way as to describe physical processes by a ``universal`` time. The different speeds of clocks are taken into account by the factor energy. The reason for that is discussed in detail in classical and quantum mechanics for freely moving masses and masses under the influence of a potential. We should consider energy as physical quantity reflecting the variation of the speed of clocks. This point of view is not treated in the literature or taught in lectures on classical mechanics, quantum mechanics and gravitation. It  could finally lead to a generalised definition of energy.

\section{Acknowledgement}
The author thanks Helmut Rauch for the interesting comment that neutron interference experiments~\cite{Rauch:1974aa} in the gravitational field can be described either by the Schrödinger equation or by the difference of proper time~\cite{Greenberger:2012aa,Scheich:2013aa}. I am also thankful to Dmitry Antonov for the critical reading of the manuscript and for indicating additional references.

\bibliographystyle{utphys}
\bibliography{literatur}

\end{document}